\documentclass[pra,aps,twocolumn,superscriptaddress,showpacs]{revtex4-1}

\usepackage{graphicx}

\begin{document}

\title{Photoelectron ionization spectra in a system interacting with a
neighbor atom}

\author{Jan Pe\v{r}ina Jr.}
\affiliation{Institute of Physics of AS CR, Joint Laboratory of
Optics, 17. listopadu 50a, 772 07 Olomouc, Czech Republic}
\author{Anton\'{i}n Luk\v{s}}
\affiliation{Palack\'{y} University, RCPTM, Joint Laboratory of
Optics, 17. listopadu 12, 771 46 Olomouc, Czech Republic}
\author{Wieslaw Leo\'nski}
\affiliation{Quantum Optics and Engineering Division, Institute of
Physics, University of Zielona G\'ora, Prof.~Z.~Szafrana 4a,
65-516 Zielona G\'ora, Poland}
\author{Vlasta Pe\v{r}inov\'{a}}
\affiliation{Palack\'{y} University, RCPTM, Joint Laboratory of
Optics, 17. listopadu 12, 771 46 Olomouc, Czech Republic}
\email{perinaj@prfnw.upol.cz}

\begin{abstract}
Photoelectron ionization spectra of a system interacting with a
neighbor two-level atom are investigated using the
Laplace-transform method. These spectra are typically composed of
several peaks. Photoelectron ionization spectra conditioned by the
measurement on the two-level atom show oscillations at the Rabi
frequency. The presence of spectral zeros occurring periodically
with the Rabi period is predicted. This phenomenon is analyzed in
detail.
\end{abstract}




\maketitle

\section{Introduction}

Ionization of an atom represents one of the most important
physical effects. In this process an electron in a bound state at
an atom is moved into a free state. There exist several mechanisms
leading to ionization, e.g., thermal ionization or ionization in a
strong static electric or magnetic field. Optical ionization is
probably the most important, because it allows a versatile
diagnostics of the electron system of the studied atom. At the
quantum level, an electron in the bound state moves to a free
state after absorbing one photon (or several photons) from the
optical field. This implies that photoelectron ionization spectra
depend strongly on the frequency of the optical field.

Shapes of the photoelectron ionization spectra are influenced by
many factors. Among them, the presence of discrete bound excited
electron states at the atom plays a dominant role as was pointed
out for the first time by Fano \cite{Fano1961}. The presence of
such excited bound states thus forms additional ionization paths
that compete with direct ionization. Along these additional paths,
an electron moves from its ground bound state to an excited bound
state due to the optical field first and, subsequently, the
Coulomb configurational interaction transfers the electron into a
free state. Interference among direct and indirect ionization
paths determines the shapes of photoelectron ionization spectra
\cite{Fano1961,Rzazewski1981,Lambropoulos1981}. Fano has shown
that there exist free states that cannot be populated owing to a
completely destructive interference of the ionization paths. This
effect is referred to as the presence of Fano zeros that do not
depend on the optical-field strength. Experimental evidence of the
Fano zeros has been brought, e.g., in \cite{Journel1993}. In
general, there exist $ N $ Fano zeros in a system with $ N $
discrete levels \cite{Fano1961}. Special attention has been
devoted to auto-ionization systems with two levels in various
configurations
\cite{Leonski1987,Leonski1988,Leonski1988a,Leonski1991}. The
interaction of auto-ionization systems with quantum optical fields
has been addressed in \cite{Leonski1990}. Also the transparency
for ultra-short optical pulses has been predicted in
auto-ionization systems \cite{Paspalakis1999}. Auto-ionization
systems can even slow-down the propagating light under certain
conditions \cite{Raczynski2006}. Similarly as other physical
effects, the dynamics of ionization can be tailored using the Zeno
and anti-Zeno effects \cite{Lewenstein2000}. Moreover, quite
recently Chu and Lin \cite{Chu2010} have proposed the method of
ultra-fast Fano resonances in the description of dynamics using
the model involving ultra-short laser pulses. It should be
stressed out that the problem of Fano resonances is not restricted
to the atomic physics only. There are numerous papers in the field
of solid state physics, especially devoted to the systems
containing quantum dots and nano-particles. For instance, the
influence of the effect of Fano resonances on the photon
statistics has been discussed in \cite{Ridolfo2010}. A wider
overview of literature devoted to the Fano resonances in
nano-physics can be found in \cite{Miroshnichenko2010}.

Here, we study the role of bound excited states located at a
neighbor atom in the formation of photoelectron ionization spectra
\cite{Luks2010}. An electron in a bound state of the neighbor atom
interacts with an electron at the ionization atom through the
dipole-dipole interaction that leads to energy transfer between
two electrons \cite{Silinsh1994}. This type of interaction (energy
transfer) qualitatively differs from that embedded in the Fano
model (electron transfer). As a consequence, the Fano zeros do not
exist in this model. However, another type of spectral zeros has
been discovered here and is named dynamical zeros. Such dynamical
zeros occur periodically with the Rabi frequency in the
conditional photoelectron ionization spectra. If bound
auto-ionizing states in the ionization system are present, both
the Fano and dynamical zeros can be found in photoelectron
ionization spectra, as will be discussed in a consecutive
publication. Suitable candidates for an experimental verification
of the presented theory are molecular condensates, in which the
electrons localized at different molecules interact through the
dipole-dipole interaction \cite{Silinsh1994}. The developed theory
can also be testified using quantum dots and other semiconductor
heterostructures \cite{Miroshnichenko2010}.

The paper is organized as follows. In Sec.~II, the model
Hamiltonian is given and the corresponding dynamical equations are
solved. Photoelectron ionization spectra in the long-time limit
are studied in Sec.~III. In Sec.~IV, the profiles of long-time
photoelectron ionization spectra are discussed and compared with
those of the Fano model. The dynamical and Fano-like spectral
zeros are investigated in Sec.~V. Conclusions are drawn in
Sec.~VI.

\section{Quantum model of the system and its dynamics}

\begin{figure}  
 \includegraphics[scale=0.7]{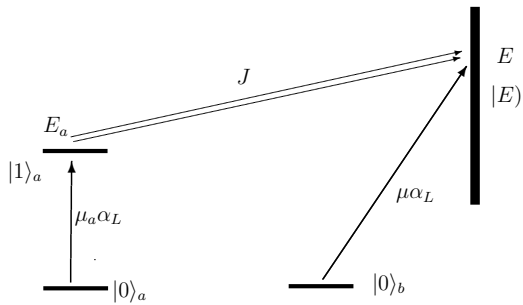}

 \caption{Scheme of the ionization system $ b $ interacting with a
  two-level atom $ a $. The state $ |1\rangle_a $ refers to an excited state
  of atom $ a $ with the energy $ E_a $ and the symbol $ |E) $ denotes
  a free state of atom $ b $ inside the continuum with the energy $ E $. The
  symbols $ \mu_a $ and $
  \mu $ stand for the dipole moments between the ground states $ |0\rangle_a $ and
  $ |0\rangle_b $ and the corresponding excited states, respectively;
  $ \alpha_L $ is the pumping amplitude.
  The constant $ J $ quantifies energy transfer due to the dipole-dipole interaction
  between the states $ |1\rangle_a $
  and $ |E) $; the double arrow indicates that two electrons at the atoms
  $ a $ and $ b $ participate in a given interaction.}
\label{fig1}
\end{figure}

We consider an ionization system that interacts with a neighbor
two-level atom by energy transfer (for the scheme, see
Fig.~\ref{fig1}).
Both atoms are under the influence of a stationary optical field.
The Hamiltonian $ \hat{H}_{\rm ion} $ of ionization atom $ b $
with flat continuum levels interacting with the optical field can
be written in the form ($ \hbar = 1 $ is assumed,
\cite{Meystre2007}):
\begin{eqnarray}   
 \hat{H}_{\rm ion} &=& \int dE  E|E)(E| \nonumber \\
 & & \hspace{-4mm} \mbox{} +
  \int dE \left[ \mu \alpha_L \exp(-iE_Lt)|E)\,{}_b\langle0|+\mbox{H.c.}
  \right] .
\label{1}
\end{eqnarray}
Here, the continuum of states of atom $ b $ is formed by the
states $ |E) $ with the energies $ E $. The dipole moment $ \mu $
characterizes an optical excitation of the continuum states; $
\alpha_L $ stands for an optical-field amplitude that oscillates
at the frequency $ E_L $. The symbol $ \mbox{H.c.} $ denotes the
Hermitian conjugate term. It should be noted that the field
frequency and the energies of atomic levels are chosen in such a
way that the possible excitation to the continuum is located far
above the ionization threshold. This assumption allows to neglect
the threshold effects and it simplifies calculations of the
integrals over the continuum energies later.

The neighbor two-level atom $ a $ with its dipole moment $ \mu_a $
being under the influence of the optical field is characterized by
the Jaynes-Cummings Hamiltonian $ \hat{H}_{\rm t-a} $:
\begin{equation}    
 \hat{H}_{\rm t-a} = E_a|1\rangle_a{}_a\langle1|
  + \left[ \mu_a \alpha_L \exp(-iE_Lt) |1\rangle_a{}_a\langle0|+\mbox{H.c.}
  \right] .
\label{2}
\end{equation}
The state $ |1\rangle_a $ refers to an excited state of atom $ a $
with the energy $ E_a $ and the ground state is denoted as $
|0\rangle_a $. We note that the ground states of atoms $ a $ and $
b $ are assumed to have the same energy that is chosen to be zero.

Energy of the interaction between the two-level atom $ a $ and the
ionization system $ b $ is given by the Hamiltonian $ \hat{H}_{\rm
trans} $:
\begin{eqnarray}   
 \hat{H}_{\rm trans} &=& \int dE
  J |E)\,{}_b\langle0||0\rangle_a{}_a\langle1|+
  \mbox{H.c.}
\label{3}
\end{eqnarray}
This energy arises from the dipole-dipole interaction between two
electrons at the atoms $ a $ and $ b $ \cite{Silinsh1994}.
Contrary to the optical dipole interaction, both electrons take
part in this interaction and change their states. One electron
returns from its excited state into the ground state, whereas the
other one takes the left energy and moves from the ground state
into its own excited state. The constant $ J $ quantifies this
energy transfer between the atoms.

We assume that the electrons at both the two-level atom $ a $ and
the ionization system $ b $ are initially in their ground states.
The interaction of two electrons at the atoms $ a $ and $ b $ with
the optical field as well as their mutual interaction lead to the
evolution that can be conveniently described in the rotating
frame. In this frame, a quantum state $ |\psi\rangle $ can be
decomposed into the following general form:
\begin{eqnarray}   
 |\psi\rangle(t) &=& c_{00}(t) |0\rangle_{a} |0\rangle_{b}
   + c_{10}(t) |1\rangle_{a} |0\rangle_{b} \nonumber \\
 & & \hspace{0mm} \mbox{}  + \int dE d_{0}(E,t)
  |0\rangle_{a} |E) \nonumber \\
 & & \hspace{0mm} \mbox{} + \int dE
  d_{1}(E,t) |1\rangle_{a} |E).
\label{4}
\end{eqnarray}
The coefficients of decomposition, $ c_{00} $, $ c_{10} $, $
d_0(E) $, and $ d_1(E) $, are time dependent.

The dynamics of the composite system governed by the
Schr\"{o}dinger equation with the Hamiltonian $ \hat{H}_{\rm ion}
+ \hat{H}_{\rm t-a} + \hat{H}_{\rm trans} $ can be conveniently
described using differential equations for the coefficients of
decomposition of the state $ |\psi\rangle $ in Eq.~(\ref{4}):
\begin{eqnarray}   
 i \frac{d}{dt} \left[\begin{array}{c}
  {\bf c}(t) \\ {\bf d}(E,t) \end{array}\right] =
  \left[\begin{array}{cc}
  {\bf A} & {\bf B}\int dE \\ {\bf B^\dagger} & {\bf K}(E)
   \end{array} \right]
  \left[\begin{array}{c} {\bf c}(t) \\ {\bf d}(E,t)
   \end{array}\right] . \nonumber \\
  & &
\label{5}
\end{eqnarray}
Here, the symbol $ \dagger $ means the Hermitian conjugation. In
Eq.~(\ref{5}), the vectors $ {\bf c} $ and $ {\bf d} $ and the
matrices $ {\bf A} $, $ {\bf B} $, and $ {\bf K} $ take the form:
\begin{eqnarray}  
 & & {\bf c}(t) = \left[\begin{array}{c}
  c_{00}(t) \\ c_{10}(t) \end{array}\right] , \hspace{5mm}
 {\bf d}(E,t) = \left[\begin{array}{c}
  d_{0}(E,t) \\ d_{1}(E,t) \end{array} \right] ,
\label{6}
  \\
 & & {\bf A} = \left[\begin{array}{cc}
  0 & \mu_a^*\alpha_L^*  \\
  \mu_a\alpha_L & \Delta E_a \end{array}\right] ,
\label{7}
  \\
 & & {\bf B} = \left[\begin{array}{cc}
   \mu^*\alpha_L^* & 0 \\ J^* & \mu^*\alpha_L^*
   \end{array}\right] ,
\label{8}
   \\
 & & {\bf K}(E) = \left[\begin{array}{cc}
  E-E_L & \mu_a^*\alpha_L^* \\
  \mu_a\alpha_L & E- E_L +\Delta E_a \end{array}\right] ;
\label{9}
\end{eqnarray}
$ \Delta E_a = E_a - E_L $. We note that the norm of state $
|\psi\rangle(t) $ is conserved during the evolution, which implies
the equality:
\begin{equation} 
 \sum_{j=0}^{1} |c_{j0}(t)|^2 + \sum_{j=0}^{1} \int dE
  \, |d_j(E,t)|^2 = 1 .
\end{equation}

The system of differential equations (\ref{5}) can be analytically
solved using the Laplace transform. Transforming Eqs.~(\ref{5}),
we arrive at the following system of algebraic equations ($
\varepsilon > 0 $):
\begin{eqnarray}   
  & & \hspace{-0.5cm} \left( \varepsilon {\bf 1_4} - \left[\begin{array}{cc}
  {\bf A} & {\bf B}\int dE \\ {\bf B^\dagger} & {\bf K}(E)
   \end{array} \right] \right)
  \left[\begin{array}{c} \tilde{\bf c}(\varepsilon) \\ \tilde{\bf d}(E,\varepsilon)
   \end{array}\right] \nonumber \\
  & & \hspace{4cm} = i \left[ \begin{array}{c} {\bf c}(0) \\ {\bf 0_2} \end{array}
   \right] ,
\label{11}
\end{eqnarray}
where the vector $ {\bf c}(0) \equiv {\bf c}(t=0) $ describes the
initial conditions. The symbol $ {\bf 1_k} $ stands for the unity
matrix in $ k $ dimensions, whereas the symbol $ {\bf 0_k} $ means
the zero vector with $ k $ elements.

The coefficients $ \tilde{d}_0(E,\varepsilon) $ and $
\tilde{d}_1(E,\varepsilon) $ describing the continuum can be
determined from the last two equations in (\ref{11}):
\begin{eqnarray}   
 \tilde{\bf d}(E,\varepsilon) = \left[ \varepsilon{\bf 1_2} -
 {\bf K}(E) \right]^{-1} {\bf B^\dagger} \tilde{\bf c}(\varepsilon) .
\label{12}
\end{eqnarray}
The substitution of Eqs.~(\ref{12}) into the first two equations
(\ref{11}) leaves us with two equations only for the coefficients
$ \tilde{c}_{00} $ and $ \tilde{c}_{10} $:
\begin{eqnarray}   
 & & \left( \varepsilon {\bf 1_2} - {\bf A} - {\bf B} \int dE
  \left[ \varepsilon{\bf 1_2} -
  {\bf K}(E) \right]^{-1} {\bf B^\dagger}  \right)
  \nonumber \\
 & & \hspace{4cm} \mbox{} \times \tilde{\bf c}(\varepsilon) = i {\bf c}(0) .
\label{13}
\end{eqnarray}

The inverse matrix $  \left[ \varepsilon{\bf 1_2} - {\bf K}(E)
\right]^{-1} $ occurring in Eq.~(\ref{13}) can be decomposed
conveniently as follows:
\begin{eqnarray}    
 \left[ \varepsilon{\bf 1_2} - {\bf K}(E) \right]^{-1} =
  \sum_{k=1}^{2} {\bf K_k} \frac{1}{
  E - \varepsilon-\xi_k} .
\label{14}
\end{eqnarray}
The energies $ \xi_1 $ and $ \xi_2 $ give oscillations of the
two-level atom $ a $:
\begin{eqnarray}   
 \xi_{1,2} &=& E_L - \frac{ \Delta E_a \pm \delta\xi }{2} , \nonumber \\
 \delta \xi &=& \sqrt{ (\Delta E_a)^2 + 4|\mu_a\alpha_L|^2 } .
\label{15}
\end{eqnarray}
The matrices $ {\bf K_1} $ and $ {\bf K_2} $ introduced in
Eq.~(\ref{14}) have the form:
\begin{equation}   
 {\bf K_k} = \frac{(-1)^k}{\delta \xi} \left[ \begin{array}{cc}
  E_a+\xi_k & -\mu_a^*\alpha_L^*  \\ -\mu_a\alpha_L &
  E_L+\xi_k \end{array} \right] .
\label{16}
\end{equation}

The integration of equation~(\ref{14}) over the energies $ E $ of
continuum from $ -\infty $ to $ \infty $ provides a useful
relation that considerably simplifies the system of equations
(\ref{13}):
\begin{equation}   
 \int dE \left[ \varepsilon{\bf 1_2} - {\bf K}(E) \right]^{-1} =
  -i\pi {\bf 1_2} .
\label{17}
\end{equation}

Equation~(\ref{13}) then represents an eigenvalue problem for a
newly defined matrix $ {\bf M} $:
\begin{eqnarray} 
 \left( \varepsilon {\bf 1_2} + {\bf M} \right) \tilde{\bf c}(\varepsilon) &=& i {\bf c}(0),
\label{18}   \\
 {\bf M} &=& {\bf A} - i\pi {\bf B} {\bf B^\dagger} .
\label{19}
\end{eqnarray}
The substitution from Eqs.~(\ref{7}) and (\ref{8}) provides the
matrix $ {\bf M} $:
\begin{eqnarray}  
 {\bf M} = \left[ \begin{array}{cc} - i\pi |\mu\alpha_L|^2 & M_{a}^c\alpha_L
  \\ M_{a}\alpha_L &  \Delta E_a - i\pi (|J|^2 + |\mu\alpha_L|^2)
  \end{array} \right] ;
\label{20}
\end{eqnarray}
$ M_{a} = \mu_a - i\pi\mu J^* $ and $ M_{a}^c = \mu_a^* - i\pi
\mu^* J $.

Introducing a matrix $ \Lambda_M $ with the eigenvalues $
\Lambda_{M,j} $ on its diagonal and the matrix of eigenvectors $
{\bf P} $ ($ {\bf M} = {\bf P}{\Lambda_M}{\bf P}^{\bf -1} $), the
solution of Eq.~(\ref{18}) can be written as follows:
\begin{equation}   
 \tilde{\bf c}(\varepsilon) = i {\bf P} {\bf U_\varepsilon} (\varepsilon)
  {\bf P}^{\bf -1} {\bf c}(0) ,
\label{21}
\end{equation}
where
\begin{equation}   
 \left[ {\bf {U_\varepsilon}} \right]_{jk}(\varepsilon) =
 \frac{\delta_{jk}}{\varepsilon - \Lambda_{M,j}}.
\label{22}
\end{equation}
The symbol $ \delta_{jk} $ means the Kronecker $ \delta $.

The inverse Laplace transform finally gives us the temporal
evolution of coefficients $ c_{00} $ and $ c_{10} $:
\begin{equation}   
 {\bf c}(t) = {\bf P} {\bf U}(t) {\bf
  P}^{\bf -1} {\bf c}(0)
\label{23}
\end{equation}
and
\begin{equation}   
 {{\bf U}}_{jk}(t) = \delta_{jk}\exp(-i\Lambda_{M,j}t) .
\label{24}
\end{equation}

The substitution of the spectral solution for the coefficients $
\tilde{\bf c} $ from Eq.~(\ref{21}) into Eq.~(\ref{12}) and the
subsequent inverse Laplace transform leave us with the temporal
coefficients $ d_0(E) $ and $ d_1(E) $ of the continuum:
\begin{eqnarray}   
 {\bf d}(E,t) &=& i \sum_{k=1}^{2} {\bf K_k} {\bf B^\dagger} {\bf P}
  {\bf U_{k}}(E,t) {\bf P}^{\bf -1} {\bf c}(0) ;
\label{25} \\
 \left[ {\bf U_{k}} \right]_{jl}(E,t) &=& \frac{i\delta_{jl}}{E-\Lambda_{M,j}-\xi_k}
   \nonumber \\
 & & \hspace{-1cm} \mbox{} \times \left[ \exp[i(\xi_k-E)t] - \exp(-i\Lambda_{M,j}t) \right] .
\label{26}
\end{eqnarray}

If we restrict our attention to the case in which both electrons
are initially in their ground states $ |0\rangle_a $ and $
|0\rangle_b $, the solution written in Eq.~(\ref{25}) can be
recast into a simplified form:
\begin{eqnarray}  
 {\bf d}(E,t) &=& i \sum_{k=1}^{2} {\bf D}_k {\bf u}_{\xi_k}(E,t) ,
\label{27} \\
 \left[{\bf u_{\xi_k}}\right]_{j}(E,t) &=& \frac{i}{E-\Lambda_{M,j}-\xi_k}
  \left[ \exp[i(\xi_k-E)t]  \right. \nonumber \\
 & &  \left. \mbox{} - \exp(-i\Lambda_{M,j}t) \right] ,
  \hspace{3mm} j=1,2.
\label{28}
\end{eqnarray}
The eigenvalues $ \Lambda_{M,j} $ of matrix $ {\bf M} $ can be
analytically derived as follows:
\begin{eqnarray}  
 \Lambda_{M,1,2} &=& {\cal E}_a - i\pi|\mu\alpha_L|^2
  \nonumber \\
 & &  \mp \sqrt{ {\cal E}_a^2 + M_{a} M_{a}^c|\alpha_L|^2 } ,
\label{29}
\end{eqnarray}
where $ {\cal E}_a = (\Delta E_a-i\gamma_a)/2 $. Moreover, the
knowledge of eigenvectors of the matrix $ {\bf M} $ allows us to
determine the matrices $ {\bf D_1} $ and $ {\bf D_2} $ introduced
in Eq.~(\ref{27}). Their matrix elements give weights to the
Lorentzian profiles that constitute the photoelectron ionization
spectra. These matrix elements $ [{\bf D_k}]_{jl} $ are defined
along the formula $ [{\bf D_k}]_{jl} = [{\bf K_k B^\dagger
P}]_{jl} [{\bf P}^{\bf -1}]_{l1} $ that provides the following
explicit relations:
\begin{eqnarray}     
 {\bf D_k} &=& \frac{1}{\delta\xi D} \left[ \begin{array}{cc}
  D_{k,11} & D_{k,12} \\
  D_{k,21} & D_{k,22} \end{array} \right] ;
\label{30}  \\
 D_{k,11} &=& [\mu \tilde{\cal E}^2 + JM_{a}\tilde{\cal E}]\alpha_L[\pm \Delta E_a/2 +
 \delta\xi/2] \nonumber \\
  & & \mbox{} \mp \mu\mu_a^* M_{a} \alpha_L |\alpha_L|^2 \tilde{\cal E} ,
  \nonumber \\
 D_{k,12} &=& [\mu M_{a} M_{a}^c \alpha_L |\alpha_L|^2 - J M_{a}\alpha_L \tilde{\cal E}]
  \nonumber \\
 & & \mbox{} \times [\pm \Delta E_a/2 + \delta\xi/2] \pm \mu\mu_a^* M_{a} \alpha_L |\alpha_L|^2
  \tilde{\cal E}, \nonumber \\
 D_{k,21} &=& \mp \mu \mu_a \alpha_L^2 \tilde{\cal E}^2 \mp J\mu_a M_{a}\alpha_L^2
  \tilde{\cal E} \nonumber \\
 & & \mbox{} + \mu M_{a} \alpha_L^2 \tilde{\cal E} [\mp \Delta E_a/2 + \delta\xi/2] ,
  \nonumber \\
 D_{k,22} &=& \mp \mu \mu_a M_{a} M_{a}^c \alpha_L^2 |\alpha_L|^2 \pm J\mu_a M_{a}\alpha_L^2
  \tilde{\cal E} \nonumber \\
 & & \mbox{} + \mu M_{a} \alpha_L^2 \tilde{\cal E} [\pm \Delta E_a/2 - \delta\xi/2] ,
  \nonumber \\
 \tilde{\cal E} &=& -{\cal E}_a - \sqrt{ {\cal E}_a^2 + M_{a}
  M_{a}^c |\alpha_L|^2 } , \nonumber \\
 D &=& -\tilde{\cal E}^2 - M_{a}M_{a}^c |\alpha_L|^2 . \nonumber
\end{eqnarray}
The Rabi frequency $ \delta \xi $ has been introduced in
Eq.~(\ref{15}).

The analysis of temporal behavior of the ionization system has
shown that the atom $ b $ is completely ionized for sufficiently
long times. The electron from the atom $ b $ is transferred to the
continuum and then left in a state that is quantum correlated
(entangled) with the state of the electron at the two-level atom $
a $. Discussion of this phenomenon will be presented elsewhere.
The photoelectron ionization spectrum is composed of four
Lorentzian curves as Eq.~(\ref{27}) indicates. Weights of these
curves change in time until they reach their long-time values. We
note that, in general, these weights depend also on the initial
conditions for the atoms $ a $ and $ b $. As we consider the
photoelectron ionization spectra obtained in stationary optical
fields, their long-time limit is of prominent interest for us.

\section{Long-time photoelectron ionization spectra}

The long-time photoelectron ionization spectral profiles are
derived from the long-time form of coefficients $ d_0(E,t) $ and $
d_1(E,t) $ given in Eq.~(\ref{25}). In this limit, i.e., for the
times $ t $ obeying $ t \gg 1/|{\rm Im}\{\Lambda_{M,j}\}| $ for $
j=1,2 $ ($ {\rm Im} $ means the imaginary part), the formulas for
the evolution matrices $ {\bf U_{k}}(E,t) $ and vectors $ {\bf
u_{\xi_k}}(E,t) $ in Eqs.~(\ref{26}) and (\ref{28}), respectively,
considerably simplify:
\begin{eqnarray}   
 \left[ {\bf U_{k}^{lt}}\right]_{jl}(E,t) &=& \delta_{jl}
 \left[{\bf u_{\xi_k}}\right]_{j}(E,t) \nonumber \\
 &=& \frac{i \delta_{jl} \exp[i(\xi_k-E)t]}{E-\Lambda_{M,j}-\xi_k} ;
\label{31}
\end{eqnarray}
the superscript $ {\rm lt} $ stands for long time. The expressions
for the evolution matrices $ {\bf U_{k}^{lt}} $ in Eq.~(\ref{31})
show that there exist two prominent frequencies $ \xi_1 $ and $
\xi_2 $ of periodic oscillations in this long-time limit. That is
why we decompose the long-time spectra $ {\bf d}^{\bf lt} $ of an
ionized electron at the atom $ b $ into parts $ {\bf d}^{\bf
\xi_1} $ and $ {\bf d}^{\bf \xi_2} $ containing oscillations at
the frequencies $ \xi_1 $ and $ \xi_2 $, respectively:
\begin{eqnarray}   
 {\bf d}^{\bf lt}(E,t) &=& {\bf d}^{\bf \xi_1}(E,t) +
  {\bf d}^{\bf \xi_2}(E,t) ,
\label{32}  \\
 {\bf d}^{\bf \xi_j}(E,t) &=& i {\bf K_{j}} {\bf B^\dagger} {\bf P}
  {\bf U_{k}^{lt}}(E,t) {\bf P}^{-1} {\bf c}(0) , \nonumber \\
  & & \mbox{} \hspace{3mm} j=1,2.
\label{33}
\end{eqnarray}
The form of the amplitude long-time photoelectron ionization
spectra $ {\bf d}^{\bf lt} $ as written in Eq~(\ref{32}) means
that the intensity photoelectron ionization spectra $ I_j^{\rm lt}
\equiv |d_j^{\rm lt}|^2 $, $ j=0,1 $, can be decomposed into two
contributions. The first steady-state contributions denoted as $
I_0^{\rm st} $ and $ I_1^{\rm st} $ are time independent, whereas
the second ones with the common magnitude $ I^{\rm osc} $
harmonically oscillate at the Rabi frequency $ \xi_1-\xi_2 =
\delta \xi $:
\begin{eqnarray}   
 I^{\bf lt}_j(E,t) &=& I^{\rm st}_j(E) + I^{\rm osc}_j(E)
  \cos[\delta\xi t+ \varphi_j(E)] , \nonumber \\
 & &  \label{34} \\
 & & {\bf \varphi}_j(E) = \arg[{\bf d}^{\bf \xi_1}_{j+1}(E)
   {\bf d}^{\bf \xi_2 *}_{j+1}(E)] , \nonumber  \\
 I^{\rm st}_j(E) &=& |{\bf d}^{\bf \xi_1}_{j+1}(E)|^2 +
  |{\bf d}^{\bf \xi_2}_{j+1}(E)|^2 ,
\label{35}  \\
 I^{\rm osc}_j(E) &=& 2 |{\bf d}^{\bf \xi_1}_{j+1}(E)|
  |{\bf d}^{\bf \xi_2}_{j+1}(E)| , \hspace{5mm} j=0,1.
\label{36}
\end{eqnarray}
The symbol $ \arg $ denotes the argument of a complex number. It
holds that $ \varphi_0(E) = \varphi_1(E) \pm \pi $ and $ I^{\rm
osc}_0(E) = I^{\rm osc}_1(E) \equiv I^{\rm osc}(E) $. Thus the
overall photoelectron ionization spectrum $ I^{\rm lt}(E) =
I_0^{\rm lt}(E,t) + I_1^{\rm lt}(E,t) $ is time independent and
can be determined along the formula:
\begin{equation}   
 I^{\rm lt}(E) = {I}^{\rm st}_0(E) + {I}^{\rm st}_1(E).
\label{37}
\end{equation}
We note that the oscillations at the Rabi frequency $ \delta \xi $
in the long-time limit can be alternatively moved into the
evolution of the state of atom $ a $ provided that a suitable
basis in the space of states of the ionized electron is chosen.

In order to observe temporal oscillations in a photoelectron
ionization spectrum, the time-resolved spectroscopy of ionized
electrons is needed. The measurement of ionization spectra has
also to be done in the post-selection regime, after the
measurement whose result guarantees the presence of an electron at
the atom $ a $ either in the ground state or the excited state.
The temporal resolution needed depends linearly on the strength of
the stationary pumping field \cite{Meystre2007}. If the
experimental temporal resolution is not sufficient, just steady
state parts $ I^{\rm st}_0(E) $ and $ I^{\rm st}_1(E) $ are
observed.

Two prominent features may characterize the photoelectron
ionization spectra in the long-time limit: the presence of the
Fano and dynamical zeros.

The term Fano zero denotes a frequency in the photoelectron
ionization spectrum that cannot be excited for any strength of the
optical pumping. The frequency $ E_F $ of a Fano zero can thus be
revealed from the condition
\begin{equation}   
 I^{\rm lt}(E_F) = 0 .
\label{38}
\end{equation}
According to Eq.~(\ref{37}), a Fano zero at the frequency $ E_F $
occurs provided that both $ I^{\rm st}_0(E_F) = 0 $ and $ I^{\rm
st}_1(E_F) = 0 $. In the investigated model, a genuine Fano zero
does not exist, since we have not included into our model an
auto-ionizing state. However, the Fano-like zeros can be observed
for the case of a weak optical pumping.

The long-time photoelectron ionization spectral intensity
components $ I^{\rm st}_0 $, $ I^{\rm st}_1 $, and $ I^{\rm osc} $
obey in general the inequalities $ I^{\rm st}_0(E) \ge I^{\rm
osc}(E) $ and $ I^{\rm st}_1(E) \ge I^{\rm osc}(E) $. This means
that there might occur the frequencies $ E_D $ fulfilling one or
both the following relations:
\begin{equation}   
 I^{\rm st}_j(E_D) = I^{\rm osc}(E_D), \hspace{5mm} j=0,1.
\label{39}
\end{equation}
At these frequencies $ E_D $, the long-time photoelectron
ionization intensity spectrum $ I_0^{\rm lt}(E,t_D) $ [or $
I_1^{\rm lt}(E,t_D) $] equals zero at specific time instants $ t_D
$. We call these frequencies $ E_D $ dynamical zeros, because they
periodically occur with the Rabi period $ 2\pi/\delta\xi $. They
represent an analogy to the usual Fano zeros in systems that
involve an additional two-level atom $ a $ oscillating in a
stationary optical field. It holds that the dynamical zeros in the
spectra $ I_0^{\rm lt} $ and $ I_1^{\rm lt} $ occur, in general,
at different frequencies $ E_D $. We will see later that the
frequencies $ E_D $ of dynamical zeros in the long-time
photoelectron spectra $ I_0^{\rm lt} $ and $ I_1^{\rm lt} $
coincide provided that the atom $ a $ is resonantly pumped.

If there exists a Fano zero at the frequency $ E_F $, there also
occurs a dynamical zero at the same frequency $ E_D = E_F $. This
can easily be understood from the conditions written in
Eqs.~(\ref{38}) and (\ref{39}) using the nonnegativity of
intensities.

\section{Discussion of shapes of long-time photoelectron ionization spectra}

There exists a similarity of the long-time photoelectron
ionization spectra in the studied ionization system interacting
with a neighbor atom and the well-known Fano model. That is why we
first summarize the main features of the spectra in the Fano model
and then we continue by comparing the spectra in both models.

\subsection{Long-time photoelectron ionization spectra in the Fano
model}

The photoelectron ionization spectra of the usual Fano model of
the atom $ b $ with one discrete excited state $ |1\rangle_b $ of
the energy $ E_b $ have been discussed, e.g., in
\cite{Fano1961,Rzazewski1981}. The state $ |1\rangle_b $ can be
optically excited through the dipole moment $ \mu_b $. It also
interacts with the continuum of states $ |E) $ by the Coulomb
configurational interaction that is characterized by a constant $
V $. Thus, there occurs competition between the direct and
indirect (through the state $ |1\rangle_b $) ionizations. The
relative strength of two ionization paths forms two distinct areas
differing in shapes of the ionization spectra.

\begin{figure}  
 (a) \includegraphics[scale=0.3]{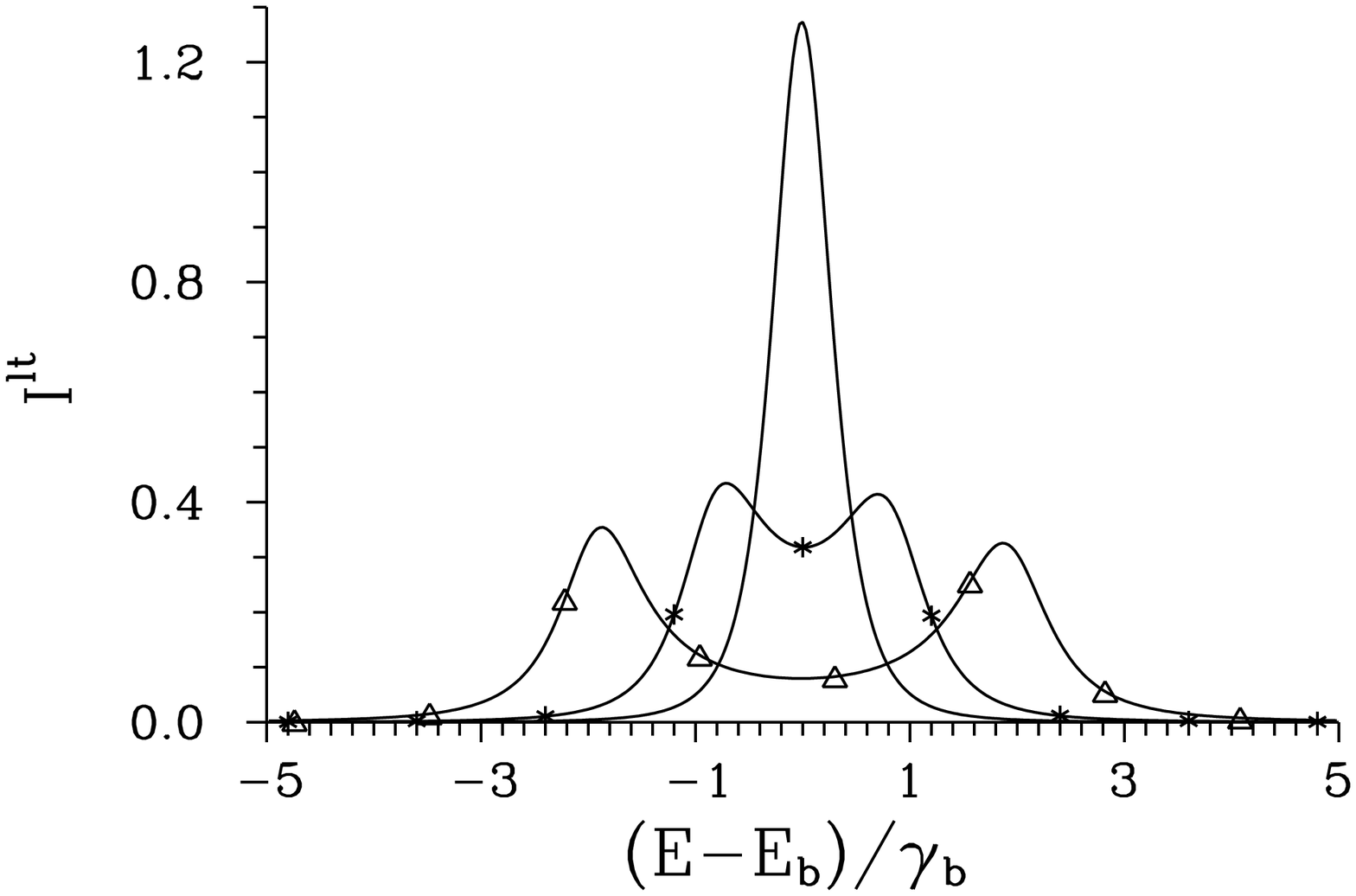}

 \vspace{7mm}
 (b) \includegraphics[scale=0.3]{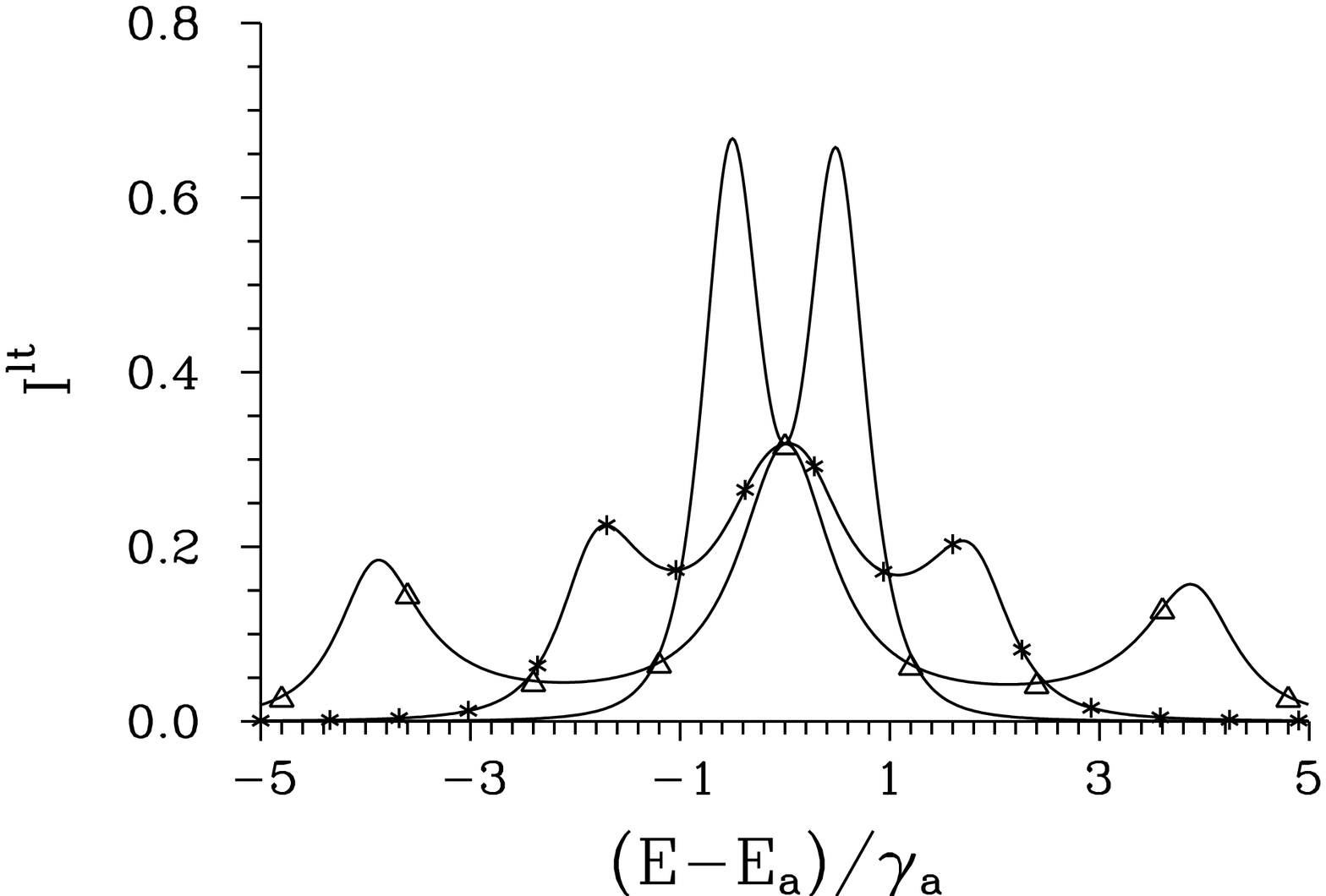}

 \caption{Long-time photoelectron ionization spectra $ I^{\rm lt} $ in
  (a) the Fano model ($ q_a = \gamma_a = 0 $, $ q_b = 100 $, $ \gamma_b = 1 $)
  and (b) the ionization system interacting with a neighbor ($ q_a = 100 $, $ \gamma_a = 1 $,
  $ q_b = \gamma_b = 0 $) in the regime of prevailing indirect ionization for
  several values of pumping parameter $ \Omega $: $ \Omega = 1
  $ (solid curve), $ \Omega = 2 $ (solid curve with $ \ast $), and
  $ \Omega = 4 $ (solid curve with $ \triangle $);
  $ q_a = \mu_a/(\pi \mu J^*) $, $ \gamma_a = \pi |J|^2 $,
  $ q_b = \mu_b/(\pi \mu V^*) $, $ \gamma_b = \pi |V|^2 $, $ \Omega =
  \sqrt{4\pi\Gamma} (Q+i)\mu\alpha_L $, $ \Gamma = \gamma_a +
  \gamma_b $, $ Q = (\gamma_a q_a + \gamma_b q_b)/\Gamma $.
  Spectra are normalized such that $ \int dE I^{\rm lt}(E)
  = 1 $; $ E_a=E_b=E_L=1 $.}
\label{fig2}
\end{figure}

We first assume the prevailing indirect optical ionization ($ q_b
= \mu_b/(\pi \mu V^*) \gg 1 $). In this case the long-time
photoelectron ionization spectrum is peaked around the frequency $
E $ being resonant with the pumping optical frequency $ E_L $ for
weaker pumping intensities [see Fig.~\ref{fig2}(a)].
Greater pump amplitudes $ \alpha_L $ lead to the Autler--Townes
splitting \cite{Autler1955} of this peak that results in a
symmetric double-peaked spectral shape. The distance between two
peaks equals $ 2|\mu_b\alpha_L| $. This means that the larger the
pump amplitude $ |\alpha_L| $, the greater the distance between
two neighbor peaks. Moreover, the widths of these peaks are
proportional to $ (\gamma_b + \pi|\mu\alpha_L|^2)/2 $ [$ \gamma_b
= \pi|V|^2 $], i.e., their broadening with the increasing pump
amplitude $ \alpha_L $ is observed. These conclusions can be drawn
from the general form of amplitude ionization spectrum that is
composed of two Lorentzian curves. These curves are centered at
the frequencies $ {\rm Re} \{\Lambda_{M,1,2}^F \} + E_L $, where $
\Lambda_{M,1,2}^F $ can be derived in the following form:
\begin{eqnarray}    
 \Lambda_{M,1,2}^F &=& \Delta E_b/2 - i\pi|\mu\alpha_L|^2/2 -
  i\gamma_b/2 \nonumber \\
 & & \hspace{-10mm} \mbox{} \mp \left[ (\Delta E_b+i\pi|\mu\alpha_L|^2 -i
  \gamma_b)^2 \right.  \nonumber \\
 & & \hspace{-10mm} \mbox{} \left.  + 4(\mu_b - i\pi\mu V^*) (\mu_b^* - i\pi\mu^* V)
  |\alpha_L|^2 \right]^{1/2} /2 ;
\label{40}
\end{eqnarray}
$ \Delta E_b = E_b - E_L $. This formula considerably simplifies
in the discussed regime if we additionally assume $ \gamma_b \leq
1 $ and $ q_b \gg 1 $:
\begin{eqnarray}    
 \Lambda_{M,1,2}^F &=& \Delta E_b/2 - i\pi|\mu\alpha_L|^2/2 -
  i\gamma_b/2 \mp |\mu_b\alpha_L|. \nonumber \\
 & &
\label{41}
\end{eqnarray}

\begin{figure}  
 (a) \includegraphics[scale=0.3]{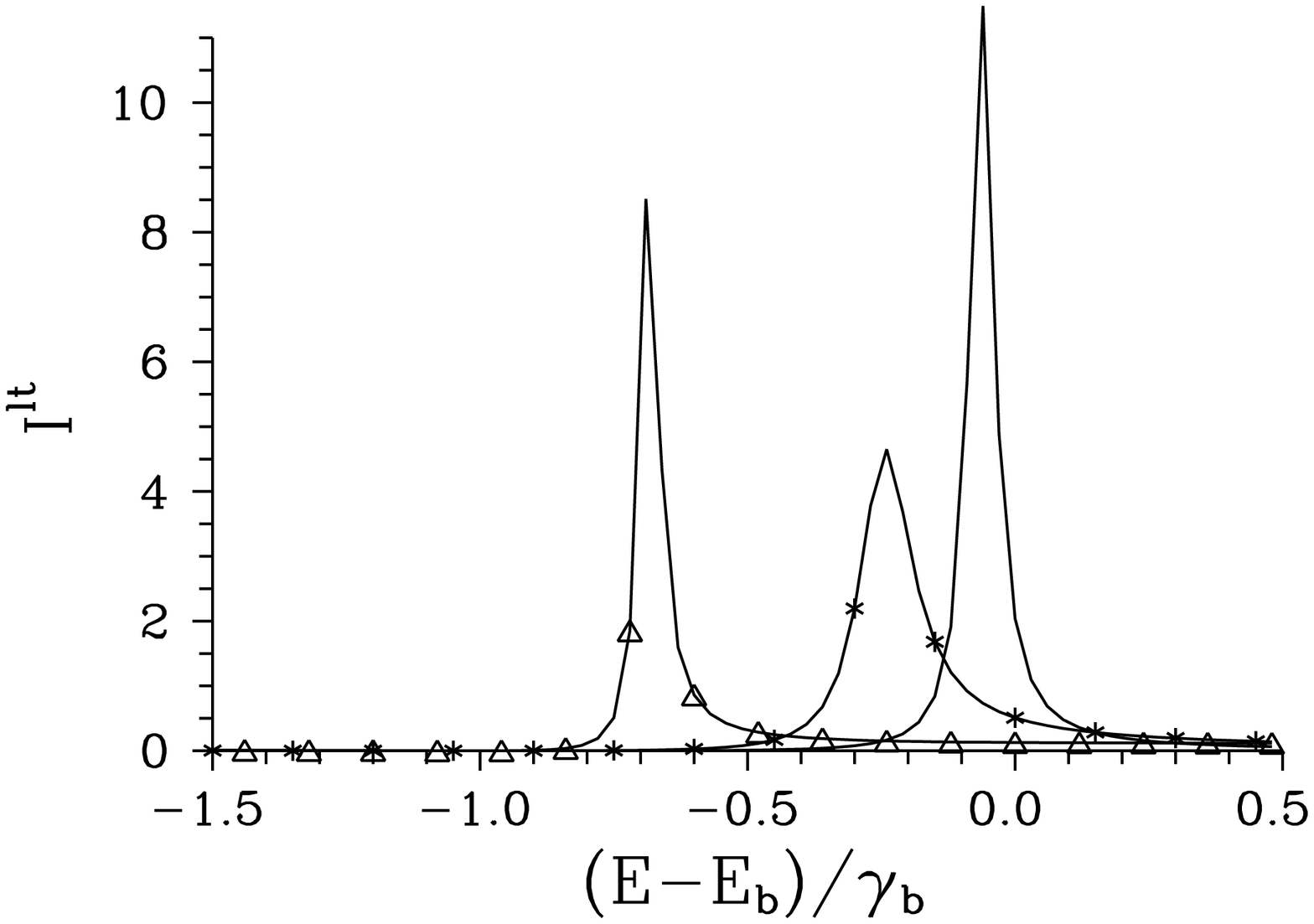}

 \vspace{7mm}
 (b) \includegraphics[scale=0.3]{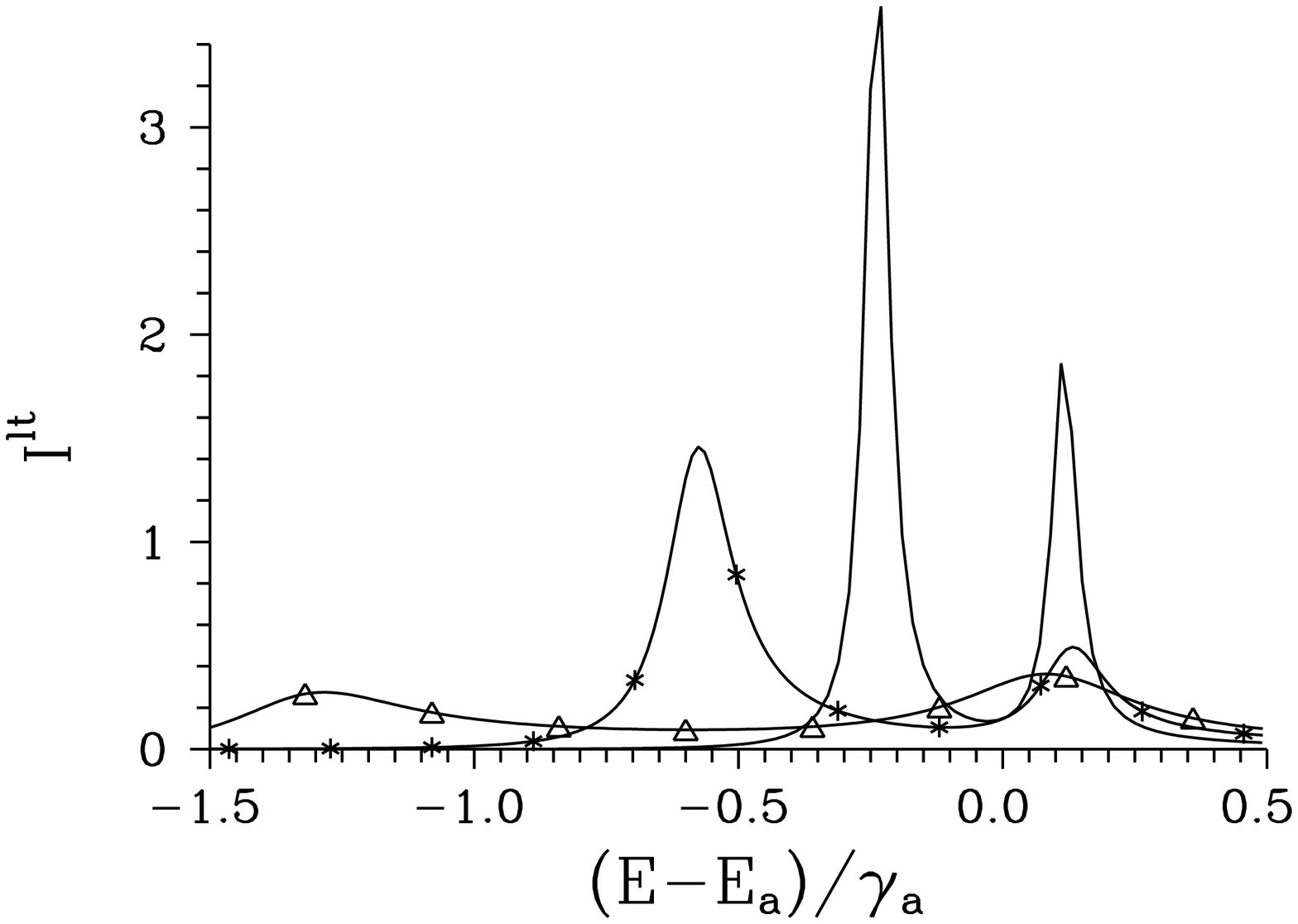}

 \caption{Long-time photoelectron ionization spectra $ I^{\rm lt} $ in
  (a) the Fano model ($ q_a = \gamma_a = 0 $, $ q_b = \gamma_b = 1 $)
  and (b) the ionization system interacting with a neighbor ($ q_a = \gamma_a = 1 $,
  $ q_b = \gamma_b = 0 $) in the regime of comparable direct and indirect
  ionization paths
  for different values of pumping parameter $ \Omega $: $ \Omega = 0.5
  $ (solid curve), $ \Omega = 1 $ (solid curve with $ \ast $), and
  $ \Omega = 2 $ (solid curve with $ \triangle $);
  $ E_a=E_b=E_L=1 $.}
\label{fig3}
\end{figure}

On the other hand, if the direct and indirect optical ionizations
are comparable, a typical shape of the photoelectron ionization
spectrum consists of one peak that moves towards the lower
frequencies $ E $ with the increasing pump amplitude $ \alpha_L $
[see Fig.~\ref{fig3}(a)]. This indicates that the Lorentzian curve
centered at the frequency $ \Lambda_{M2}^F $ given in
Eq.~(\ref{40}) is decisive for the spectral shape. In general,
there occurs one Fano zero that originates in the destructive
interference of two ionization paths. The frequency $ E_F $ of
this zero is expressed as $ E_F = E_b - \gamma_b q_b $.

\subsection{Long-time photoelectron ionization spectra of an ionization
system interacting with a neighbor}

Now we pay attention to photoelectron ionization spectra of the
ionization system interacting with a two-level neighbor atom.
Here, in parallel to the direct ionization, the ionization of atom
$ b $ may also occur after the energy transfer from the excited
bound state $ |1\rangle_a $ of atom $ a $, that is not accompanied
by the electron transfer. This indirect ionization path interferes
with the path of direct ionization of the atom $ b $. The presence
of the second electron at the atom $ a $, that undergoes the
long-time stationary Rabi oscillations, substantially modifies the
photoelectron ionization spectra of the Fano model discussed
above. As a consequence, the photoelectron ionization spectra $
d_0^{\rm lt}(E,t) $ and $ d_1^{\rm lt}(E,t) $ belonging to the
atom $ a $ in the ground and excited states, respectively, can be
decomposed into two contributions oscillating at the prominent
frequencies $ \xi_1 $ and $ \xi_2 $ [see Eq.~(\ref{15})]. The
overall photoelectron ionization spectrum $ I^{\rm lt} $ defined
in Eq.~(\ref{37}) is constituted by four Lorentzian curves
centered around the frequencies $ {\rm Re} \{\Lambda_{M,j} \}
+\xi_k $ for $ j,k=1,2 $ (the symbol $ {\rm Re} $ denotes the real
part of an argument).

As we want to compare both models at comparable conditions, we
assume that the strengths of interactions leading to
auto-ionization are comparable, i.e. $ \gamma_a \approx \gamma_b
$. The special case of $ \gamma_a \ll \gamma_b $ appropriate for
molecular condensates with their dipole-dipole and Coulomb
configurational interactions is discussed at the end of Sec.~IV.

Provided that the indirect ionization is much stronger than the
direct one (for this case, the Fano-like asymmetry parameter $ q_a
= \mu_a/(\pi \mu J) \gg 1 $), the assumptions of resonant pumping
($ E_a = E_L) $ and the weaker dipole-dipole interaction ($
\gamma_a \leq 1 $, $ q_a \gg 1 $) allows us to express four
complex central frequencies in a simple form:
\begin{eqnarray}  
 \Lambda_{M,1,2} + \xi_{1,2} &=& E_L - i\gamma_a/2
  - i\pi|\mu\alpha_L|^2 \nonumber \\
 & & \mbox{} \mp |\mu_a\alpha_L| \mp |\mu_a\alpha_L|.
\label{42}
\end{eqnarray}
These formulas indicate that the long-time photoelectron
ionization spectrum consists of one central peak and two symmetric
side-peaks shifted by the frequency $ \pm 2 |\mu_a\alpha_L| $. It
can be shown that two side-peaks dominate the photoelectron
spectrum for weak optical pumping, whereas three distinct peaks
can be observed for stronger pumping [see Fig.~\ref{fig2}(b)]. The
widths of the spectral peaks are proportional to $ \gamma_a/2 +
\pi|\mu\alpha_L|^2 $, i.e., the greater the pump amplitude $
\alpha_L $ the broader the peaks. Compared to the long-time
photoelectron ionization spectra of the Fano model, we
additionally have the central peak and also the mutual distance of
two side-peaks is twice that found in the Fano model [see
Eq.~(\ref{42})]. For greater pump amplitudes $ \alpha_L $ ($
\gamma_a \ll |\mu\alpha_L|^2 $), the widths of the spectral peaks
are twice compared to those of the Fano model.

If the direct and indirect ionizations are comparable ($ q_a
\approx 1 $), two peaks dominate the long-time photoelectron
ionization spectrum, as demonstrated in Fig.~\ref{fig3}(b). The
first peak is located near the pumping frequency $ E_L $, whereas
the second one occurs around the frequency $ E_L -
2|\mu_a\alpha_L| $. The second peak thus moves towards the lower
frequencies when increasing the pump amplitude $ \alpha_L $. This
behavior is qualitatively similar to that found in the Fano model
[see Fig.~\ref{fig3}(a)]. However, the frequency shift is two
times larger in the ionization system interacting with a neighbor.
Moreover, the spectral peaks are two times broader for stronger
pumping.

\begin{figure}  
 \includegraphics[scale=0.3]{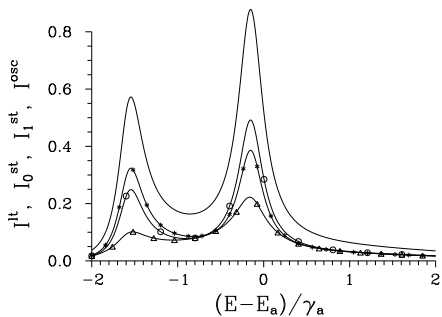}

 \caption{Long-time photoelectron ionization spectrum $ I^{\rm lt} $ (solid curve), its
  steady-state components $ I^{\rm st}_0 $ (solid curve with $ \circ $) and $ I^{\rm st}_1 $
  (solid curve with $ \ast $), and magnitude  $ I^{\rm osc} $ of the harmonically oscillating part
  (solid curve with $ \triangle $) for the ionization system interacting
  with a neighbor under non-resonant pumping;
  $ q_a = \gamma_a = 1 $, $ q_b = \gamma_b = 0 $,
  $ \Omega = 2 $, $ E_a=1 $, $ E_L=0.8 $.}
\label{fig4}
\end{figure}

Contrary to the Fano model, the long-time photoelectron spectrum $
I^{\rm lt} $ can be decomposed into two parts $ I^{\rm lt}_0 $ and
$ I^{\rm lt}_1 $ that are conditioned by the presence of the
electron at the two-level neighbor atom $ a $ in the states $
|0\rangle_a $ and $ |1\rangle_a $, respectively. The conditional
spectra $ I^{\rm lt}_j $ are described by their steady-state
components $ I^{\rm st}_j $ and the common term $ I^{\rm osc} $
that harmonically oscillates at the Rabi frequency $ \delta\xi $.
If the neighbor atom $ a $ is resonantly pumped, $ I^{\rm st}_0 =
I^{\rm st}_1 $. Otherwise, the two steady-state components differ
in their profiles. This is illustrated in Fig.~\ref{fig4}. For
this case, two dynamical zeros around the relative frequency $
(E-E_a)/\gamma_a = - 0.6 $ appear in the spectrum. Indeed, a
closer inspection of the curves in Fig.~\ref{fig4} reveals that,
for two frequencies in this area, the amplitude $ I^{\rm osc} $ of
the oscillating part is equal to the value of either the
steady-state part $ I^{\rm st}_0 $ or the steady-state part $
I^{\rm st}_1 $. Moreover, it should be mentioned that, the
occurrence times of the two dynamical zeros in the profiles of
spectra $ I^{\rm lt}_0 $ and $ I^{\rm lt}_1 $ are mutually shifted
by half the Rabi period $ \pi/\delta\xi $.

\begin{figure}  
 \includegraphics[scale=0.3]{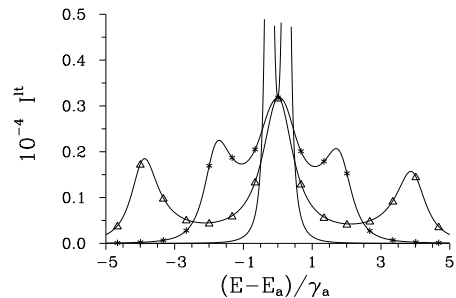}
 \caption{Long-time photoelectron ionization spectra $ I^{\rm lt} $ in
  the ionization system interacting with a neighbor for different values
  of pumping parameter $ \Omega $: $ \Omega = 5 \times 10^{-5}
  $ (solid curve), $ \Omega = 1 \times 10^{-4} $ (solid curve with $ \ast $), and
  $ \Omega = 5 \times 10^{-4} $ (solid curve with $ \triangle $);
  $ q_a = 100 $, $ \gamma_a = 1 \times 10^{-4} $,
  $ q_b = \gamma_b = 0 $, $ E_a=E_b=E_L=1 $.}
\label{fig5}
\end{figure}

If molecular condensates are considered, $ \gamma_a \ll \gamma_b
$. In molecular condensates, the Coulomb configurational
interaction constant $ V $ is typically in eV, whereas the
dipole-dipole interaction (hopping) constant $ J $ is of the order
of $ 1 - 10 $~meV \cite{Silinsh1994}. This implies that $
\gamma_a/\gamma_b \approx 10^{-4} - 10^{-6} $. Due to the scaling
properties of the model the spectral profiles obtained in the
regime $ \gamma_a \approx \gamma_b $ can also be observed in this
case provided that the optical pumping described by parameter $
\Omega $ is weaker by four or six orders of magnitude. As an
example, the long-time photoelectron ionization spectra determined
for $ \gamma_a = 1 \times 10^{-4} $ are shown in Fig.~\ref{fig5}.
Their profiles are similar to those valid for $ \gamma_a = 1 $ and
drawn in Fig.~\ref{fig2}(b). However, we should emphasize that,
for $ \gamma_a \ll 1 $, the case of comparable dipole moments $
\mu_a $ and $ \mu $ of atoms $ a $ and $ b $, respectively, is
characterized by the condition $ q_a \gg 1 $. On the other hand,
if $ q_a \approx 1 $ then $ \mu_a \ll \mu $. Small values of
dipole moments $ \mu_a $ of the neighbor atom $ a $ can be
reached, e.g., by the detuning of atom $ a $ from the resonance
with the pumping field applying a dc electric field.

\section{Dynamical and Fano-like zeros}

The frequencies of dynamical and Fano-like (for a weak optical
pumping) zeros are distinguished features of the long-time
photoelectron ionization spectra. We can even find them
analytically in the limit of weak optical pumping ($ \alpha_L
\rightarrow 0 $) using perturbation expansions for the matrices $
{\bf D_j} $ in Eq.~(\ref{30}) and the frequencies $ \Lambda_{M,j}
$ in Eq.~(\ref{29}) and $ \xi_j $ in Eq.~(\ref{15}).

\subsection{Fano-like zeros}

We first pay attention to a weak optical pumping that is in
resonance with the two-level neighbor atom $ a $ ($ E_a=E_L $).
Under this condition, it is sufficient to express the matrices $
{\bf D_{\bf 1,2}} $ in the first power of pump amplitude $
\alpha_L $ and the frequencies in the zeroth power of $ \alpha_L
$. We thus obtain:
\begin{eqnarray} 
 & & \hspace{-10mm} {\bf D_{\bf 1,2}} = \left[ \begin{array}{cc} -\frac{iJ M_{a}}{\gamma_a}
 & -\mu + \frac{iJ M_{a}}{\gamma_a}
  \\  \pm \frac{iJ M_{a}}{\gamma_a} \frac{\mu_a}{|\mu_a|} &
  \pm \left( \mu - \frac{iJ M_{a}}{\gamma_a} \right)
  \frac{\mu_a}{|\mu_a|} \end{array}
   \right] \frac{\alpha_L}{2} ,
\label{43} \\
 & & \hspace{-10mm} \Lambda_{M,1,2} + \xi_j = E_L - i\gamma_a/2 \mp i\gamma_a/2 ,
  \hspace{3mm} j=1,2.
\label{44}
\end{eqnarray}

As the rows of the matrices $ {\bf D_1} $ and $ {\bf D_2} $ in
Eq.~(\ref{43}) are the same up to a multiplicative complex factor,
we obtain the same equations for the frequencies of the Fano-like
zeros both in the spectra $ d_0^{\rm lt} $ and $ d_1^{\rm lt} $ as
well as in their spectral component oscillating at the frequencies
$ \xi_1 $ and $ \xi_2 $. These equations together guarantee the
fulfilment of the condition for the Fano zero given in
Eq.~(\ref{38}). They have the common form:
\begin{equation}   
  \frac{iJ M_{a}}{\gamma_a (E_{F-l}-E_L+i\gamma_a)} +
  \frac{\mu-iJ M_{a}/\gamma_a }{E_{F-l}-E_L} = 0.
\label{45}
\end{equation}
The solution of Eq.~(\ref{45}) gives the frequency $ E_{F-l} $ of
Fano-like zero at the position:
\begin{equation}  
 \frac{E_{F-l} -E_a}{\gamma_a} = - q_a ;
\label{46}
\end{equation}
$ q_a = \mu_a/(\pi\mu J^*) $.

In the case of weak optical pumping that is out-of-resonance with
the two-level atom $ a $ ($ E_a \neq E_L $), we discuss possible
positions of the Fano-like zeros separately for the components
oscillating at the frequencies $ \xi_1 $ and $ \xi_2 $. As for the
frequency $ \xi_1 $, terms in the first power of $ \alpha_L $ in
the spectrum $ d_0^{\rm lt} $ as well as terms in the third power
of $ \alpha_L $ in the spectrum $ d_1^{\rm lt} $ lead to the same
equation. We note that the restriction to the first power of $
\alpha_L $ in the expressions for the frequencies $ \Lambda_{M,j}
$ and $ \xi_j $ is sufficient in this derivation. The solution of
the resulting equation is independent of the detuning $ \Delta E_a
$ and coincides with that derived for the resonant pumping in
Eq.~(\ref{46}).

On considering the frequency $ \xi_2 $, the general formula for
the matrices $ {\bf D}_j $ in Eq.~(\ref{30}) can be approximated
by terms written in the second power of $ \alpha_L $ both for the
spectra $ d_0^{\rm lt} $ and $ d_1^{\rm lt} $. These terms form
two different equations for the frequencies $ E_{F-l} $ of
Fano-like zeros. However, the solutions of both equations coincide
and we have:
\begin{equation}  
 \frac{E_{F-l} -E_a}{\gamma_a} = - q_a - \frac{2\Delta E_a}{\gamma_a} +
  \frac{i\Delta E_a}{q_a \gamma_a} .
\label{47}
\end{equation}
Equation (\ref{47}) indicates that the non-resonant pumping of
atom $ a $ moves the energy $ E_{F-l} $ of a possible Fano-like
zero into the complex plane $ E $. This means that there cannot
occur any Fano-like zero in this case.

The Fano zeros in the ionization system interacting with a
neighbor have been looked for numerically in the regime of
stronger optical pumping. However, no Fano zero has been revealed.

\subsection{Dynamical zeros}

The frequencies $ E_D $ of dynamical zeros in the limit of weak
optical pumping have been analyzed analytically in the same vein.
The numerical analysis has then been found useful for arbitrarily
strong pumping. The frequencies $ E_D $ of dynamical zeros are
given by the general formula in Eq.~(\ref{39}). Equivalently, this
formula can be replaced by a more suitable one:
\begin{equation}   
 |d_j^{\xi_1}(E,t)| = |d_j^{\xi_2}(E,t)|, \hspace{5mm} j=0,1.
\label{48}
\end{equation}
Similarly as in the case of Fano-like zeros, a separate discussion
of the resonant and non-resonant pumping is convenient.

On assuming a weak optical pumping that is resonant with the atom
$ a $ ($ E_a = E_L $), the Taylor expansions of the matrices ${\bf
D_1} $ and ${\bf D_2} $ in Eq.~(\ref{30}) to the second power of
pump amplitude $ \alpha_L $ together with the Taylor expansion of
frequencies $ \Lambda_{M,j} $ and $ \xi_j $ to the first power of
$ \alpha_L $ result in the following equation for the frequencies
$ E_D $:
\begin{eqnarray}    
 && \hspace{-1cm} \left| \frac{ J M_{a} -\mu \mu_a^* M_{a} \alpha_L /
  |\mu_a| }{ E_D - E_a +i\gamma_a - |\mu_a\alpha_L| } \right. \nonumber \\
 && \left. +
  \frac{ -i\mu \gamma_a - J M_{a} + \mu \mu_a^* M_{a} \alpha_L /
  |\mu_a| }{ E_D - E_a - |\mu_a\alpha_L| } \right|  \nonumber
  \\
 && \hspace{-1cm}  = \left| \frac{ -J M_{a} -\mu \mu_a^* M_{a} \alpha_L /
  |\mu_a| }{ E_D - E_a +i\gamma_a + |\mu_a\alpha_L| } \right. \nonumber \\
 && \left. +
  \frac{ i\mu \gamma_a + J M_{a} + \mu \mu_a^* M_{a} \alpha_L /
  |\mu_a| }{ E_D - E_a + |\mu_a\alpha_L| } \right| .
\label{49}
\end{eqnarray}
We note that this equation is valid for both the spectra $
d_0^{\rm lt} $ and $ d_1^{\rm lt} $, i.e., the frequencies of
dynamical zeros in these spectra coincide.

Equation (\ref{49}) can be recast into a polynomial of the fifth
order in the normalized frequency $ \bar{E} $, $ \bar{E} =
(E-E_L)/\gamma_a $:
\begin{eqnarray}   
 && |p_a|^2{\rm Im}\{p_a\} \bar{E}^5 + [- 2{\rm Re}\{p_a\} + {\rm
  Im}\{p_a^2\}] \bar{E}^4 \nonumber \\
 && \hspace{5mm}  \mbox{} + [|p_a|^2 + |p_a|^2{\rm
  Im}\{p_a\} - 2] \bar{E}^3 \nonumber \\
 && \hspace{10mm} \mbox{} + {\rm Im}\{p_a^2\} \bar{E}^2 - \bar{E} =
  0;
\label{50}
\end{eqnarray}
$ p_a = 1/q_a $. This means that up to five dynamical zeros can
occur, in principle, at the same positions in the spectra $
I_0^{\rm lt} $ and $ I_1^{\rm lt} $. One root of the polynomial in
Eq.~(\ref{50}) equals zero. Provided that the parameter $ q_a $ is
real, the remaining fourth-order polynomial collapses into the
third-order polynomial:
\begin{eqnarray}   
 \bar{E}^3 + \left(q_a - \frac{1}{2q_a} \right) \bar{E}^2 +
  \frac{q_a}{2} = 0 .
\label{51}
\end{eqnarray}
The direct inspection gives one root of this polynomial at the
normalized frequency $ \bar{E} = - q_a $. This frequency coincides
with the frequency of Fano-like zero given in Eq.~(\ref{46}). The
remaining two roots can then be recovered easily. Thus, we arrive
at the following four normalized frequencies of dynamical zeros:
\begin{eqnarray}   
 \frac{E_D-E_a}{\gamma_a} &=& 0 , \nonumber \\
 \frac{E_D-E_a}{\gamma_a} &=& - q_a , \nonumber \\
 \frac{E_D-E_a}{\gamma_a} &=& \frac{1}{4q_a} \pm \frac{1}{4}
  \sqrt{ \frac{1}{q_a^2} -8 } .
\label{52}
\end{eqnarray}
The last two normalized frequencies written in Eq.~(\ref{52}) are
real provided that $ |q_a| \le 1/(2\sqrt{2}) $.

\begin{figure}  
 (a) \includegraphics[scale=0.3]{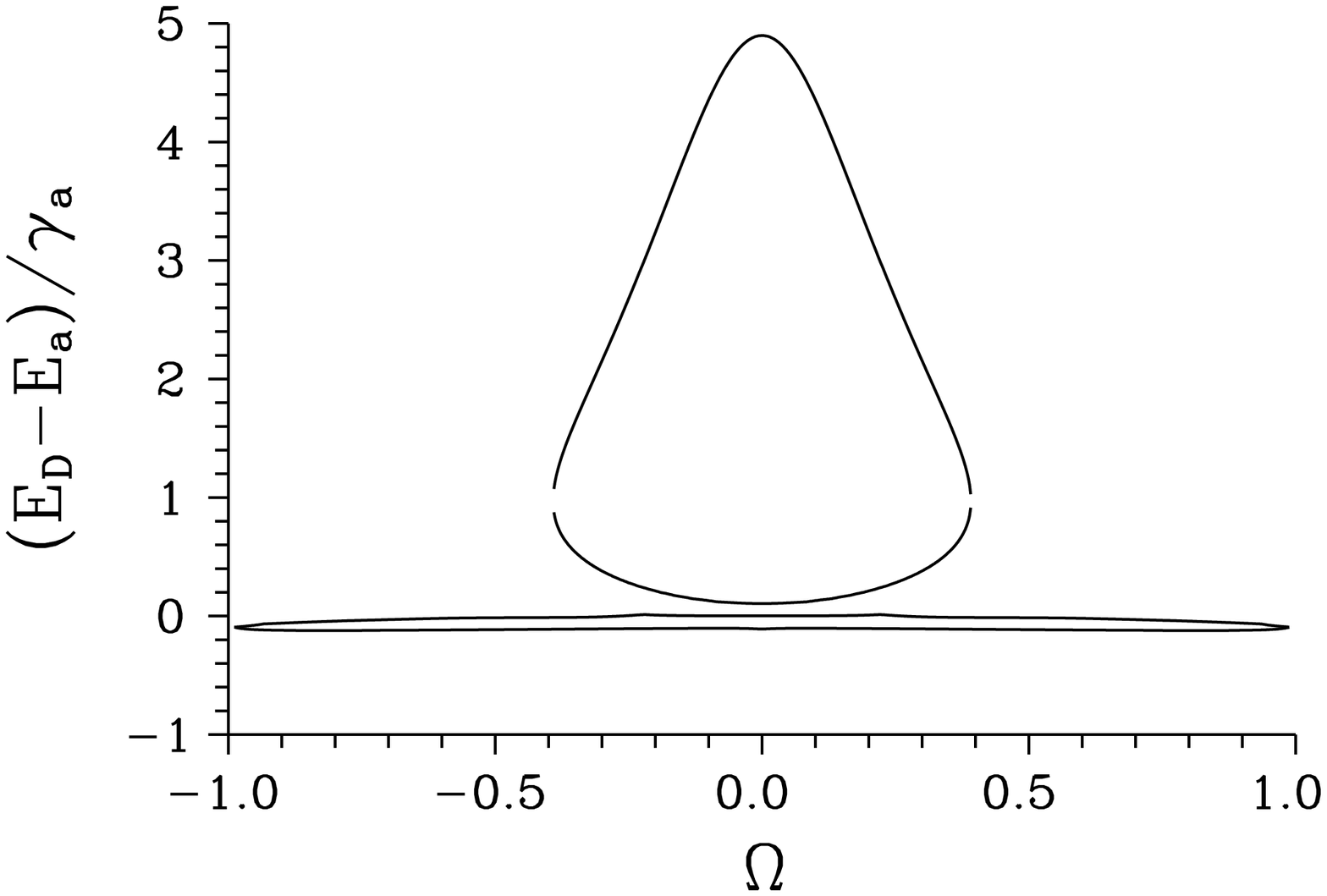}

 \vspace{7mm}
 (b) \includegraphics[scale=0.3]{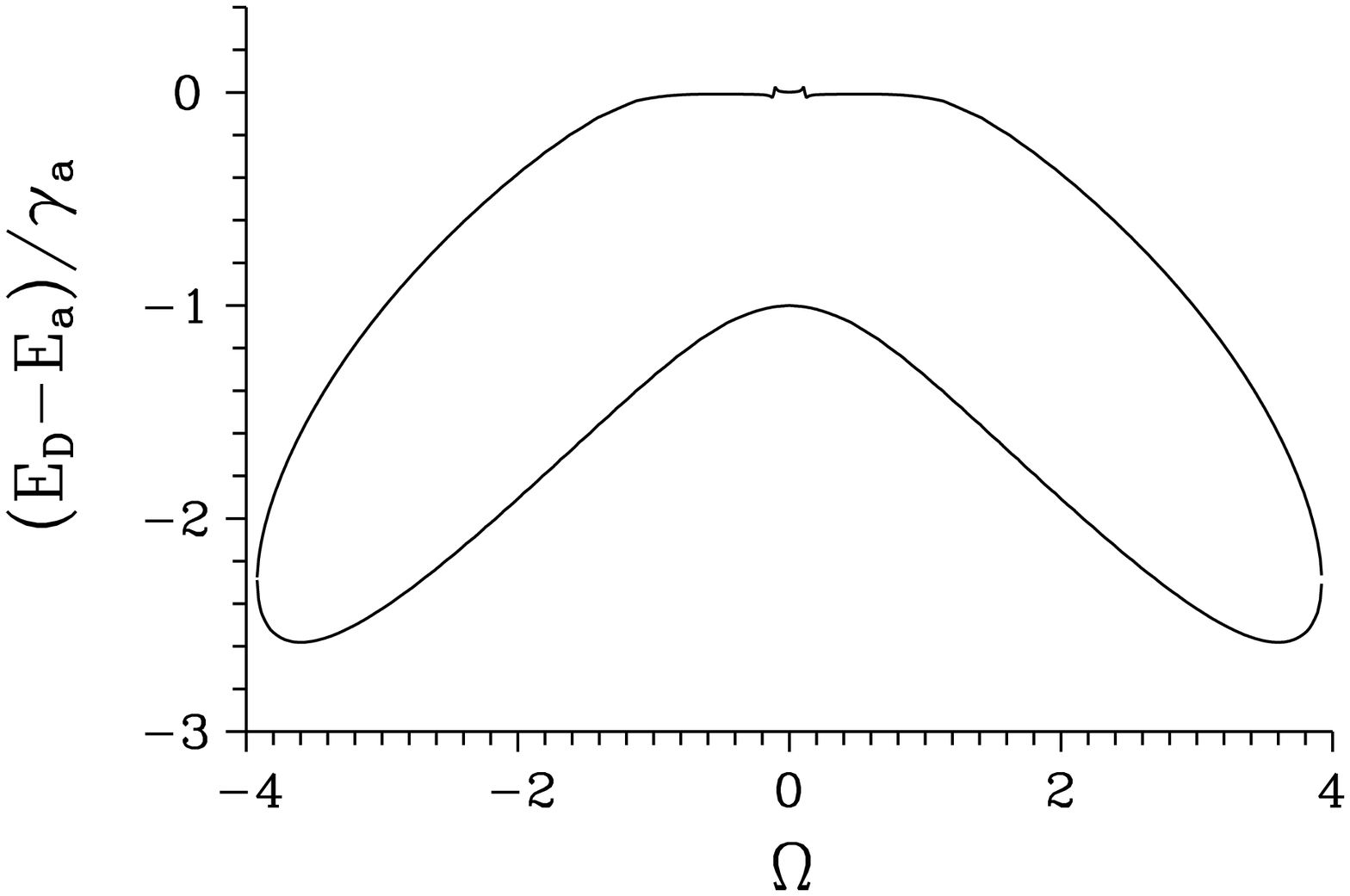}

 \vspace{7mm}
 (c) \includegraphics[scale=0.3]{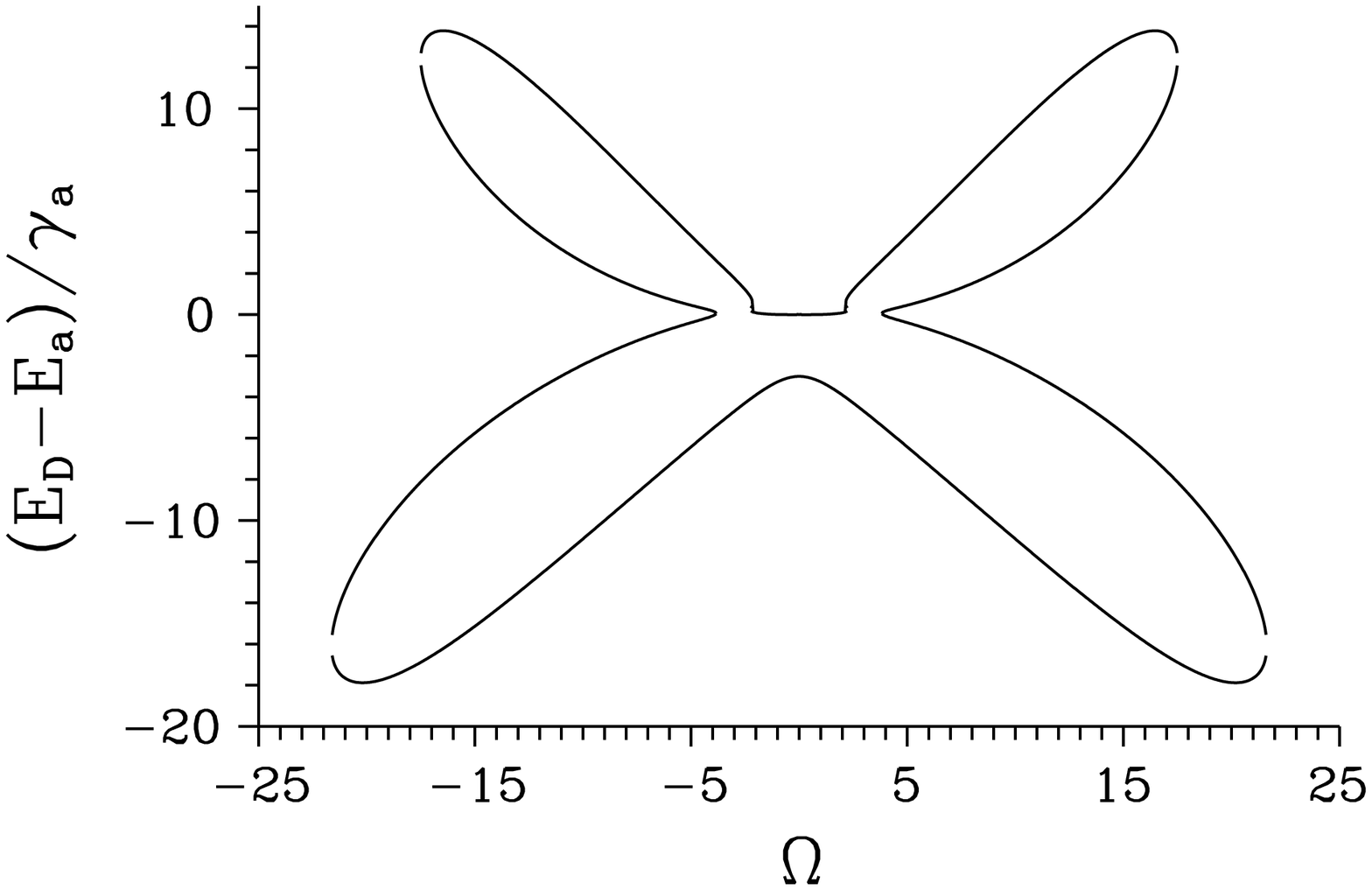}

 \caption{Normalized frequencies $ (E_D-E_a)/\gamma_a $ of dynamical zeros
  as they depend on pumping parameter $ \Omega $ for different values
  of parameter $ q_a $: (a) $ q_a = 0.1 $, (b) $ q_a = 1 $, and
  (c) $ q_a = 3 $;
  $ \gamma_a = 1 $, $ q_b = \gamma_b = 0 $, $ E_a=E_L=1 $.}
\label{fig6}
\end{figure}

In the case of arbitrarily strong optical pumping, up to five
dynamical zeros can be discovered. They usually occur in pairs.
When studying the dependence of frequencies $ E_D $ on the pumping
strength $ \Omega $ (for the definition, see the caption to
Fig.~\ref{fig2}, $ \Omega \propto \alpha_L $), the creation and
annihilation of frequency pairs has been observed (see
Fig.~\ref{fig6}). We have found three qualitatively different
shapes of curves in graphs showing the dependence of normalized
frequencies $ (E_D-E_a)/\gamma_a $ on the pumping parameter $
\Omega $ depending on the relative strengths of direct and
indirect ionization paths. They are plotted in Fig.~\ref{fig6}. In
Fig.~\ref{fig6}, the values of the frequencies $ E_D $ for $
\Omega \rightarrow 0 $ coincide with those analytically described
in Eq.~(\ref{52}). We also note that the graphs in Fig.~\ref{fig6}
are symmetric with respect to the exchange $ \Omega $ by $ -\Omega
$.

On considering the non-resonant pumping of atom $ a $ ($ E_a \neq
E_L $), the frequencies $ E_D $ of dynamical zeros, in general,
differ for the photoelectron ionization spectra $ I_0^{\rm lt} $
and  $ I_1^{\rm lt} $. On assuming a weak optical pumping $
\alpha_L $ and the spectrum $ I_0^{\rm lt} $, the Taylor expansion
of the matrices $ {\bf D_j} $, $ j=1,2 $, in Eq.~(\ref{30}) up to
the first power in $ \alpha_L $ is sufficient. The condition for
dynamical zeros in Eq.~(\ref{48}) then reveals two normalized
frequencies $ (E_D-E_a)/\gamma_a $:
\begin{eqnarray}  
 \frac{E_D-E_a}{\gamma_a} &=& - q_a , \nonumber \\
 \frac{E_D-E_a}{\gamma_a} &=& - 2\frac{\Delta E_a}{\gamma_a} .
\label{53}
\end{eqnarray}

On the other hand, the Taylor expansion of matrices $ {\bf D_j} $,
$ j=1,2 $, up to the second power in $ \alpha_L $ is needed when
studying the frequencies $ E_D $ in the photoelectron ionization
spectrum $ I_1^{\rm lt} $. The condition for dynamical zeros
written in Eq.~(\ref{48}) can be recast into a fifth-order
polynomial in the frequencies $ E_D $. On assuming the real
parameter $ q_a $, this polynomial reduces to the fourth-order
polynomial in the normalized frequency $ \bar{E} =
(E_D-E_L)/\gamma_a $:
\begin{eqnarray}   
 && \hspace{-3mm} \left[ 4q_a + \frac{1}{q_a^2} \delta E_a \right]
  \bar{E}^4  + \left[ -2 + 4q_a^2 + \frac{2}{q_a^2}
  \delta E_a^2 \right] \bar{E}^3 \nonumber \\
 && \hspace{0mm} \mbox{} - \left[ \left( 3+ \frac{1}{q_a^2} \right) \delta
  E_a + 4q_a \delta E_a^2 + \frac{1}{q_a^2} \delta E_a^3 \right]
  \bar{E}^2 \nonumber \\
 && \hspace{0mm} \mbox{} + \left[ 2q_a^2 - 2q_a\delta E_a \right]
   \bar{E} + \left[ q_a^2 \delta E_a - 2q_a\delta E_a^2 + \delta E_a^3
   \right] = 0, \nonumber \\
 &&
\label{54}
\end{eqnarray}
where $ \delta E_a = \Delta E_a / \gamma_a $. We note that $
\Delta E_a \neq 0 $ was assumed in the derivation of
Eq.~(\ref{54}). There exist, in general, four roots of this
polynomial that can be found numerically. As only real roots give
the frequencies $ E_D $, there might occur 0, 2, or 4 frequencies
$ E_D $ determining positions of dynamical zeros.

\begin{figure}  
 \vspace{5mm}

 (a) \includegraphics[scale=0.3]{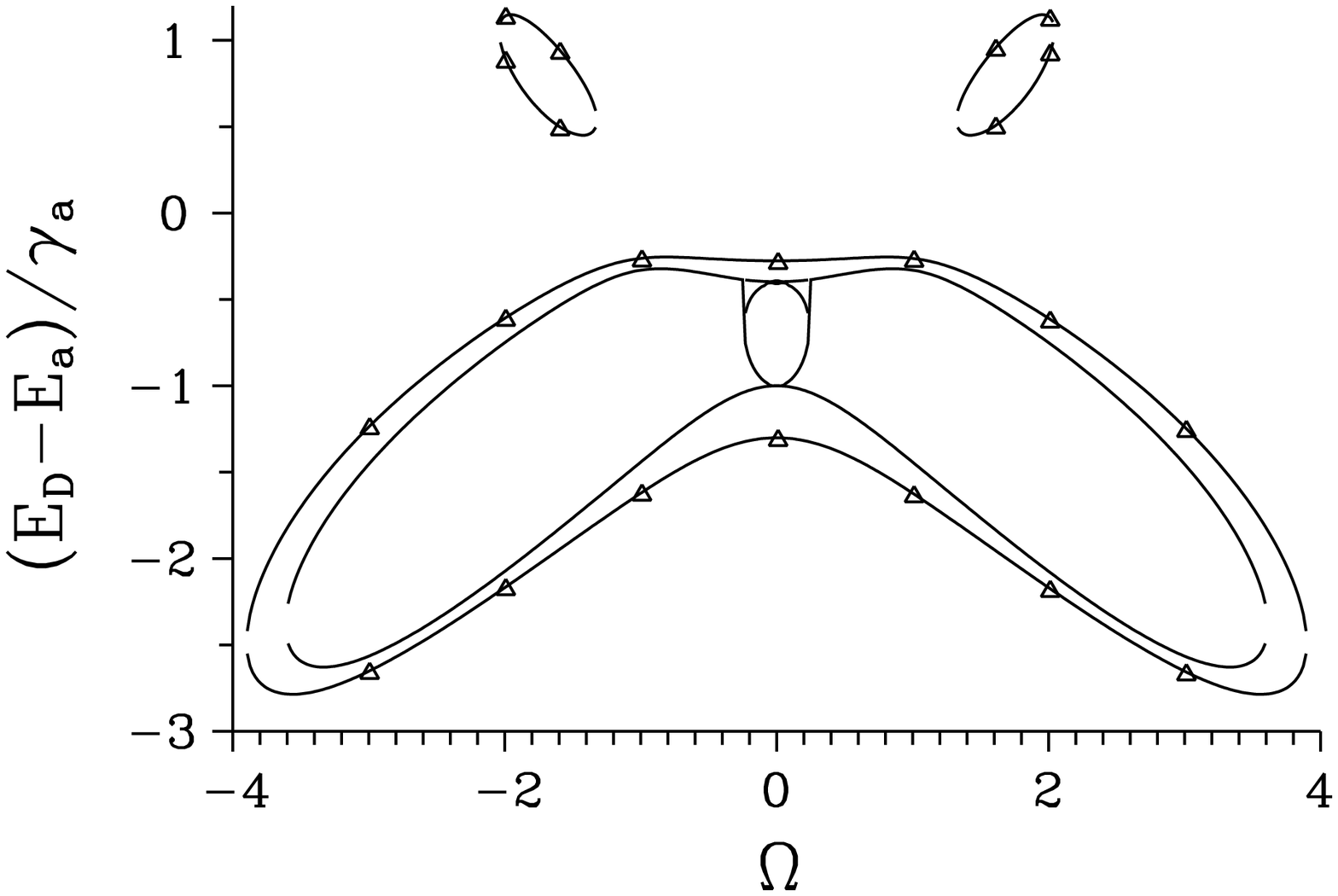}

 \vspace{7mm}
 (b) \includegraphics[scale=0.3]{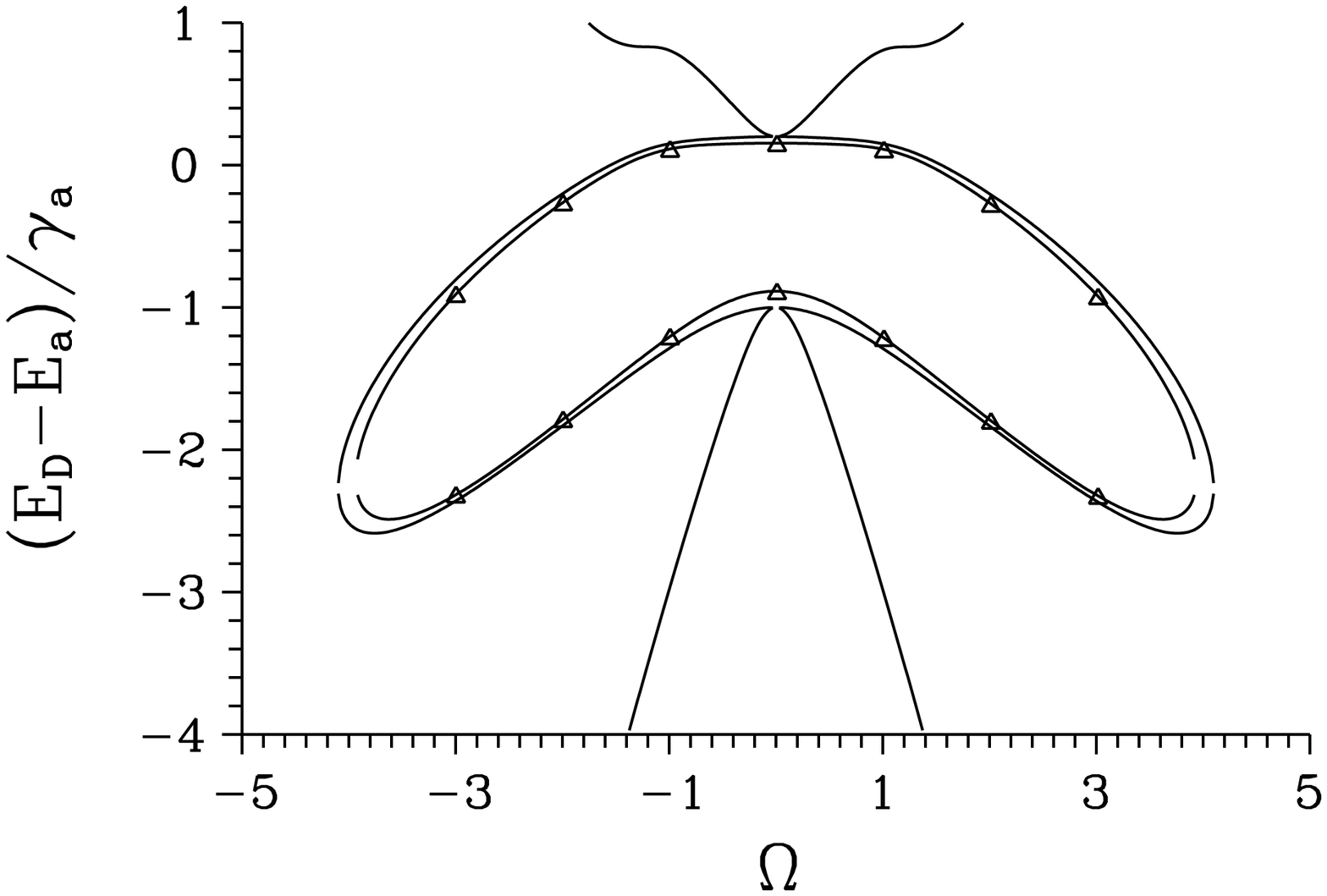}

 \caption{Normalized frequencies $ (E_D-E_a)/\gamma_a $ of dynamical zeros
  as they depend on pumping parameter $ \Omega $ for the non-resonant pumping of atom $ a $:
  (a) $ E_L = 0.8 $ and (b) $ E_L = 1.1 $.
  Solid curves indicate the frequencies found in the spectra
  $ I^{\rm lt}_0 $, solid curves with $ \triangle $
  correspond to the spectra $ I^{\rm lt}_1 $;
  $ q_a = \gamma_a = 1 $, $ q_b = \gamma_b = 0 $, $ E_a=1 $.}
\label{fig7}
\end{figure}

Typical behavior of the dynamical zeros for the non-resonant
pumping of atom $ a $ considered as a function of the pumping
parameter $ \Omega $ is shown in Fig.~\ref{fig7} for $ q_a = 1 $.
The comparison with the graph in Fig.~\ref{fig6}(b) reveals two
characteristic features caused by the non-resonant pumping. First,
the curves observed for the resonant pumping are split into two
adjacent ones: one belongs to the normalized frequencies of
dynamical zeros in the spectra $ I^{\rm lt}_0 $, one gives the
normalized frequencies in the spectra $ I^{\rm lt}_1 $. Second,
there occur new curves with shapes depending qualitatively on the
sign of the detuning $ \Delta E_a $. The symmetry $ \Omega
\leftrightarrow -\Omega $ is also preserved.

We would like to emphasize a prominent role of parameter $
\gamma_a $ in the determination of frequencies $ E_D $ of the
dynamical zeros. The parameter $ \gamma_a $ giving the strength of
'non-optical' auto-ionization interaction (e.g., dipole-dipole
interaction) occurs in the calculations only as a scaling
parameter of the frequency difference $ E_D-E_a $. This means that
the obtained results are applicable, after appropriate rescaling,
also to the case of molecular condensates in which the
dipole-dipole interaction gives $ \gamma_a \approx 10^{-4} -
10^{-6} $.

\section{Conclusions}

The long-time photoelectron ionization spectra of an ionization
system interacting with a neighbor two-level atom have been
investigated using the Laplace-transform method. They have been
compared with the spectra characterizing the Fano model of
auto-ionization. The spectra are typically composed of several
peaks of different widths depending on the pump-field intensity.
As a consequence of the interference of two ionization paths,
zeros in the long-time photoelectron ionization spectra may occur.
Whereas the genuine Fano zeros cannot be found in this model, the
Fano-like zeros occurring for a weak optical pumping can be
observed. The long-time photoelectron ionization spectra
conditioned by the presence of the neighbor two-level atom in a
given state exhibit the permanent Rabi oscillations. Dynamical
spectral zeros observed once in the Rabi period have been
discovered in these conditional spectra. The frequencies of
dynamical zeros depend on the strength of optical pumping as well
as on the projected state of the two-level atom. The numbers as
well as the values of frequencies of the dynamical zeros have been
analyzed in detail. Molecular condensates have been shown to be
suitable candidates for the experimental verification of the
predicted effect.

\acknowledgments Support by the projects 1M06002, COST OC 09026,
and Operational Program Research and Development for Innovations -
European Social Fund (project CZ.1.05/2.1.00/03.0058) of the
Ministry of Education of the Czech Republic as well as the project
IAA100100713 of GA AV \v{C}R is acknowledged.

\bibliography{perina}
\bibliographystyle{apsrev}

\end{document}